\documentclass[12pt]{spieman}  
\usepackage{amsmath,amsfonts,amssymb}
\usepackage{graphicx}
\usepackage{setspace}
\usepackage{tocloft}
\usepackage{siunitx}
\usepackage{soul}

\title{Investigation of Deferred Charge Effects in LSST ITL Sensors}

\author[a, b, c, *]{Adam Snyder}
\author[b, c]{Aaron Roodman}

\affil[a]{Stanford University, Department of Physics, 450 Serra Mall, Stanford, CA, USA}
\affil[b]{Kavli Institute for Particle Astrophysics and Cosmology, Stanford University, Stanford, CA, USA}
\affil[c]{SLAC National Accelerator Laboratory, 2575 Sand Hill Road, Menlo Park, CA, USA}

\cftpagenumbersoff{figure}
\cftpagenumbersoff{table} 
\begin{document} 
\maketitle

\begin{abstract}
The traditional characterization of charge transfer inefficiency (CTI) in charge-coupled devices (CCDs) can suffer from a number of deficiencies: CTI is often only calculated for a limited number of signal levels, CTI is calculated from a limited number of pixels, and the sources of CTI are usually assumed to occur at every pixel-to-pixel transfer.  A number of serial CTI effects have been identified during preliminary testing of CCDs developed by Imaging Technology Laboratory (ITL) for use in the Large Synoptic Survey Telescope (LSST) camera focal plane that motivate additional study beyond the traditional CTI characterization.  This study describes a more detailed examination of the serial deferred charge effects in order to fully characterize the deferred charge measured in the serial overscan pixels of these sensors.  The results indicate that in addition to proportional CTI loss that occurs at each pixel transfer, ITL CCDs have additional contributions to the deferred charge measured in serial overscan pixels, likely caused by fixed CTI loss due to charge trapping, and an electronic offset drift at high signal.

\end{abstract}


{\noindent \footnotesize\textbf{*}Adam Snyder,  \linkable{snyder18@stanford.edu} }

\begin{spacing}{1.25}   

\section{Introduction}
\label{sect:intro}  

\begin{figure}[htb]
\begin{center}
\begin{tabular}{c}
\includegraphics[height=8cm]{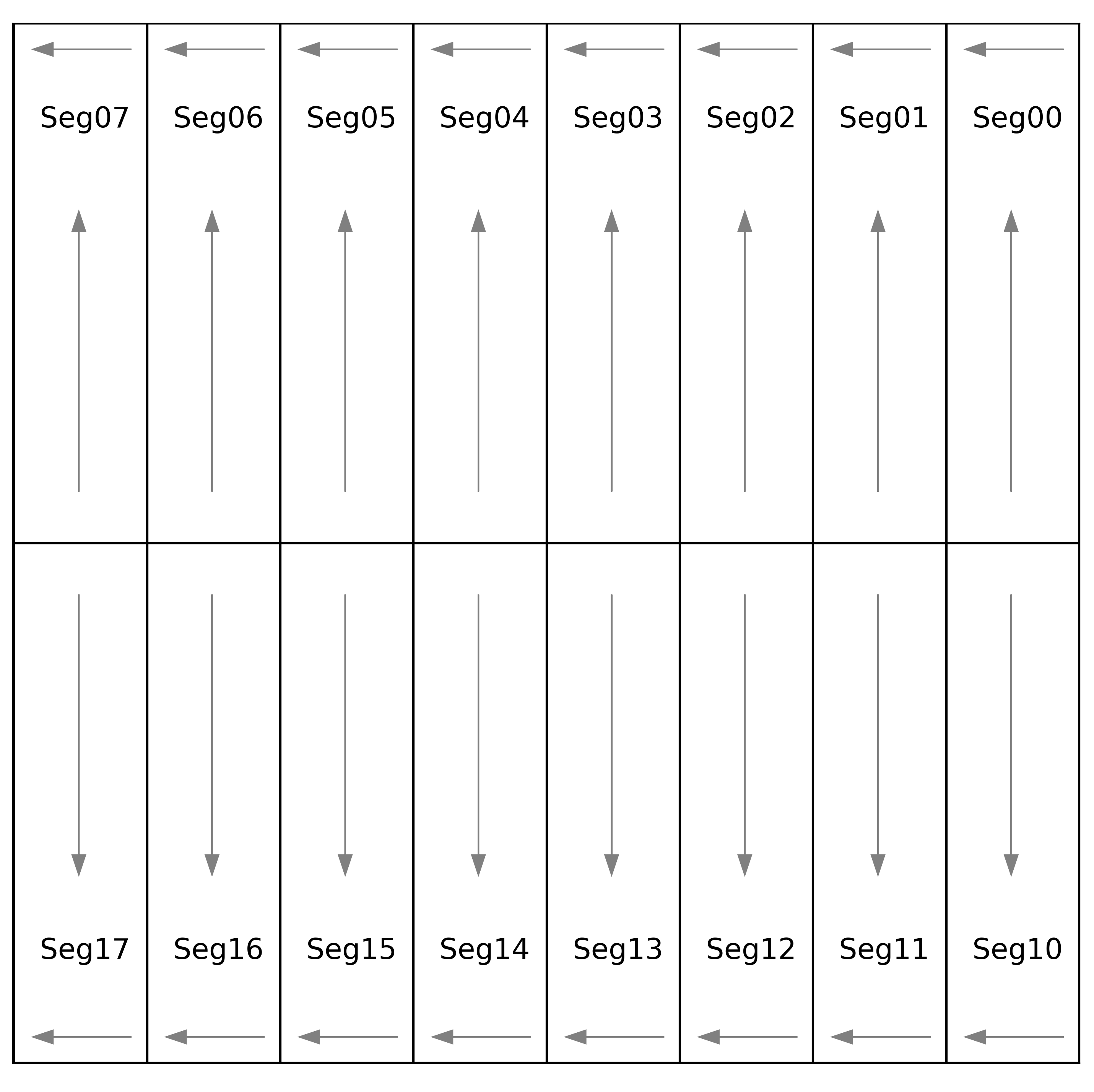}
\end{tabular}
\end{center}
\caption 
{ \label{fig:itl_layout}
LSST ITL CCD segment layout with segment name labels.  Arrows indicate the direction of parallel and serial transfer during readout.} 
\end{figure}

Integration and testing of the Large Synoptic Survey Telescope (LSST) camera sub-components is currently being performed at the SLAC National Accelerator Laboratory\cite{Bond2018, Ivezic2008}.  The camera focal plane consists of 189 charge-coupled devices (CCDs) arranged into 21 stand-alone Raft Tower Modules (RTMs) that include the necessary electronic, mechanical, and thermal support components \cite{Kahn2010}.   Each individual 4000 x 4072 CCD is sub-divided into 16 separate readout amplifier channels, in order to meet the 2 second readout time requirement (Figure \ref{fig:itl_layout}).  One of the most important CCD electro-optical properties that is measured during device characterization is the determination of the amount of charge that does not fully transfer pixel-to-pixel during readout.  The incomplete transfer of charge results in a spurious trail of signal (most noticeable in bright sources) that can affect precision measurements of source position and shape\cite{Massey2009}.

The amount of deferred charge that occurs during readout is most often characterized by the charge transfer inefficiency (CTI), defined as the ratio of electrons not transferred between two neighboring pixels, to the total electrons before the transfer, and is measured for both parallel and serial pixel transfers.  The CTI is calculated (Equation \ref{eq:eper}) from the ratio of signal in the overscan pixels $S_O$ compared to the signal in the last imaging pixel $S_{LP}$, corrected for the total number of serial transfers $N_T$, known as the extended pixel edge response (EPER) method \cite{Janesick2001}.

\begin{equation}
\label{eq:eper}
CTI = \frac{S_O}{S_{LP} N_T}
\end{equation}

The LSST specification for serial CTI and parallel CTI is less than $5 \times 10^{-6}$ and less than $3 \times 10^{-6}$ respectively, calculated from flat field images taken at signals of 1,000 electrons and 50,000 electrons (compared to the pixel full well requirement of less than 175,000 electrons) \cite{OConnor2016}.  Results of the initial CTI measurements indicated that several segments on CCDs manufactured by Imaging Technology Laboratory (ITL) failed to meet the serial CTI specification for low signal (1,000 electrons), motivating a more detailed analysis of serial deferred charge for these CCDs, over a larger range of signals, that is presented in this work.

\section{Serial CTI measurements from EPER}
\label{sect:serialcti}

The data used in this study was taken with a science-grade RTM consisting of ITL 3800C CCDs cooled to \ang{-100} C and operated under voltage and clocking conditions currently planned for on-sky imaging.  The most relevant operating voltages, in regards to serial charge transfer, are the serial clock high and low voltages (5.0 V and -5.0 V respectively) and the output gate voltage (-2.0 V); a more detailed study on the effect of ITL CCD operating voltage conditions is described in a previous study by Snyder, Gilmore, and Roodman\cite{Snyder2018}).   The full frame readout time was 2 seconds, or 2 $\mu$seconds per single pixel readout.  Serial voltage wave shaping optimization had been previously performed during single sensor testing.  Some limited image data, taken under sensor operating temperatures of \ang{-90} C was also used for comparison.  The results presented here are from a single sensor that is representative of a typical ITL sensor; all LSST ITL sensors show similar high and low signal serial deferred charge effects.  

Serial CTI was calculated from a series of calibrated flat field images (originally acquired for linearity and photon transfer gain measurements) with signal levels ranging from 100 electrons to 150,000 electrons (85\% of full well), where blooming full well effects begin to become dominant.  Electronic offset correction of the images was performed on a row-by-row basis, using the mean value of the serial overscan pixels (64 overscan pixels per row), omitting the first 5 and last 2 overscan pixels.  The omission of the first 5 serial overscan pixels was done to remove the contributions from deferred charge; results presented in this paper show this to be insufficient and in need of modification for future sensor analyses.  The omission of the last 2 serial overscan pixels is a legacy from earlier single sensor testing where there was substantial noise in the final serial overscan pixels.  A bias frame correction using an offset-corrected superbias image was also performed, in order to remove any additional electronic spatial non-uniformity.  CTI calculations from EPER were made using the signal in the first two overscan pixels.  Photon shot noise and read noise (less than 7 electrons) are greatly suppressed in the following measurements of serial deferred charge by averaging over the results for each of the individual 2000 pixel array rows.

\begin{figure}[htb]
\begin{center}
\begin{tabular}{c}
\includegraphics[height=7.5cm]{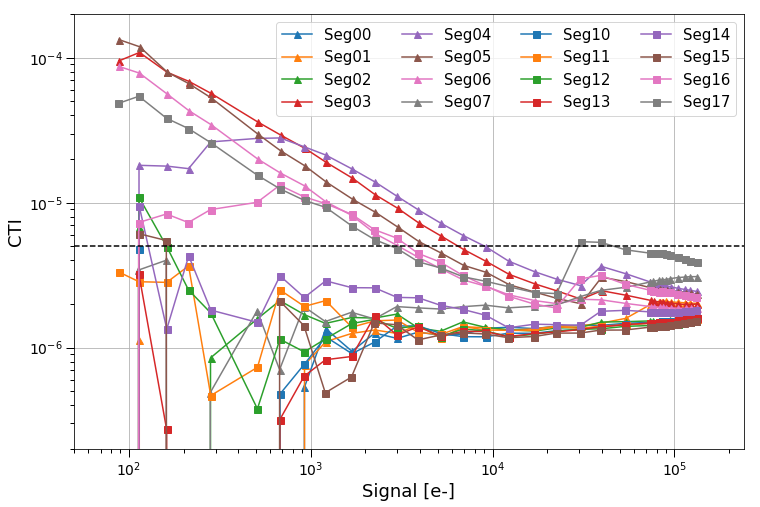}
\end{tabular}
\end{center}
\caption 
{ \label{fig:cti}
Serial CTI calculated using the EPER method for the 16 segments of a single ITL 3800C CCD.  A subset of segments exhibit large CTI at low signals.} 
\end{figure}

Measurements of the CTI using the EPER method, as a function of flat field signal (Figure \ref{fig:cti}) show two distinct populations of segments; those that exhibit anomalous large CTI at low signal (hereinafter referred to as "low signal inefficient" segments), and those that exhibit near constant CTI across the full well signal range.  A subgroup of low signal inefficient segments show a roll-off of CTI values at the lowest measured signal levels.

LSST CCDs are manufactured from high quality silicon and not subject to highly energetic particles due to radiation.  Thus they are expected to be close to "bulk-state-limited" and have an extremely small number of radiation-induced traps.  Deferred charge at each pixel transfer may be introduced due to the fast sensor readout, due to the interplay of the characteristic time scales for self-induced drift, fringing field drift and thermal drift compared to the pixel transfer time.  For these reasons, the CTI is assumed to be dominated by proportional loss (deferred charge proportional to signal) that occurs at every pixel transfer and have a weak dependence on signal.  The results presented in Figure \ref{fig:cti} challenge both of these assumptions.  A more detailed study of the overscan pixel signal values for low signal levels is presented in Section \ref{sect:lowsignal}.  A parallel study of the overscan pixel signal values for high signal levels showed an additional deferred charge effect that does not appear in CTI from EPER measurements.  These results are presented in Section \ref{sect:highsignal}.

\section{Low Signal Deferred Charge}
\label{sect:lowsignal}

\begin{figure}[htb]
\begin{center}
\begin{tabular}{c}
\includegraphics[height=10cm]{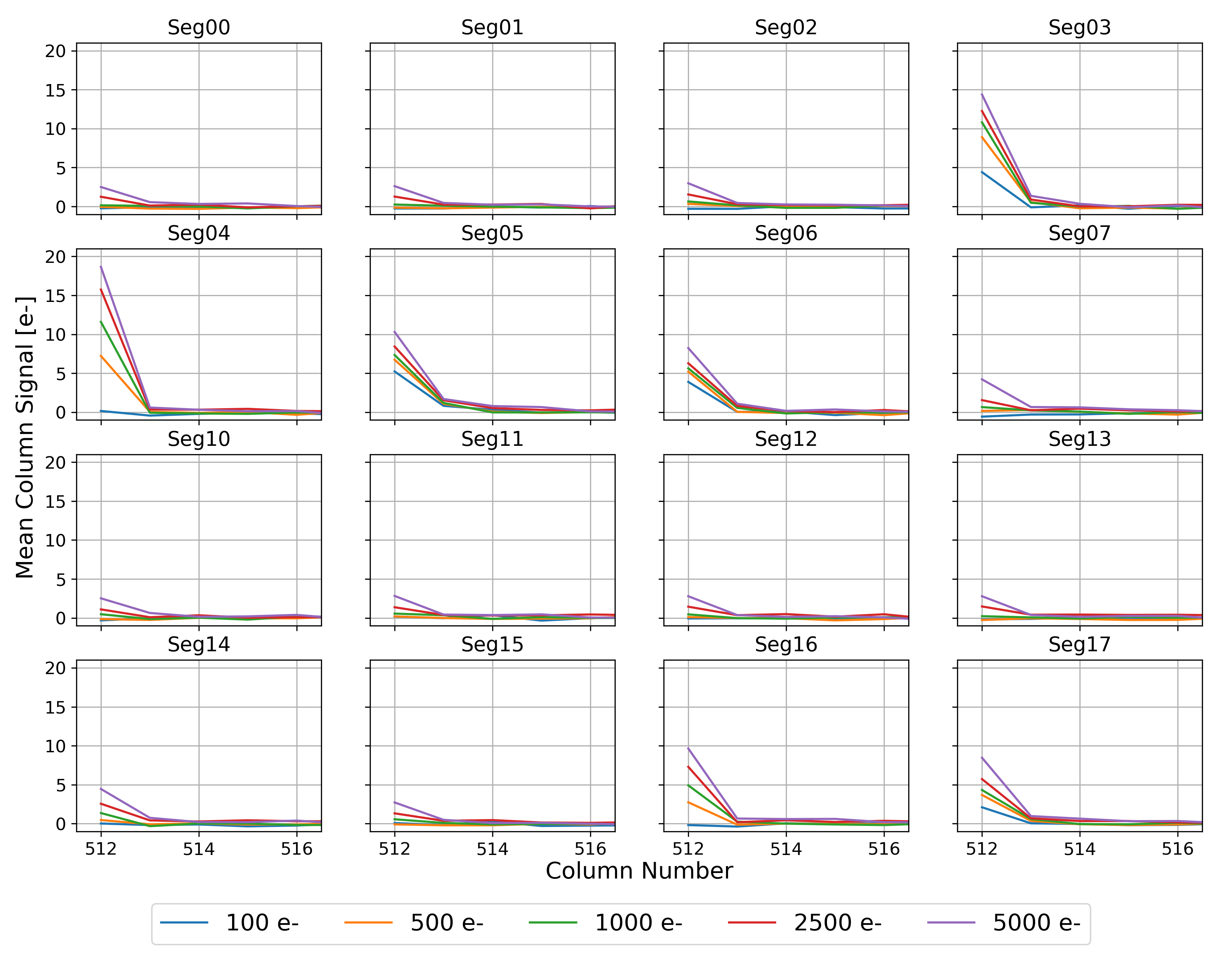}
\end{tabular}
\end{center}
\caption 
{ \label{fig:eper_low}
Mean signal in overscan columns after a flat field image (on a linear scale), for five different low signal levels, for each CCD segment.  High deferred charge signal is observed in a subset of segments (Seg03, Seg04, Seg05, Seg06, Seg16, Seg17). Error on the mean column signal is 0.14 electrons.}
\end{figure}

The deferred charge in overscan pixels following flat field images can be measured by calculating the mean signal along an overscan column, as a function of column number.  The results for a subset of flat field image signal levels below 5000 electrons, where the low signal inefficient segments exhibit measured CTI values greater than the LSST specification, are shown in Figure \ref{fig:eper_low}.  The deferred charge in the first and second overscan pixels for the low signal inefficient segments is much larger than the nominal segments and shows an exponential tail.

 The behavior of the CTI curves as a function of signal and an exponential tail in the serial overscan pixels, for low signal inefficient segments, can be explained as being caused by additional serial deferred charge due to trapping of O(10) electrons during serial readout.  For most of the low signal inefficient segments, saturation of this trapping appears to occur at signals below 100 electrons, resulting in fixed loss and decreasing CTI as a function of signal.  The roll-off of the CTI in Seg04 and Seg16 at the lowest signal levels would indicate that the trapping in these segments does not saturate until signal levels of several hundred electrons, resulting in proportional loss due to trapping until saturation is reached.
 
Exponential tail fitting of the overscan pixels can be performed by approximating the overscan pixel signal $S_O$ as a function of overscan pixel number $x$ and flat field signal $S_F$ by a contribution from proportional loss $\mathrm{CTI}_P$ occurring at every pixel transfer, accumulated over $N_T$ transfers, plus the exponential release of a fixed amount of charge $C$ with emission time constant $\tau$ (Equation \ref{eq:lowflux}).

\begin{equation}\label{eq:lowflux}
S_O(x, S_F) = C e^{-x/\tau} + \mathrm{CTI_P}^x N_T S_F.
\end{equation} 

A per segment best fit for the parameters $C$, $\tau$ and $\mathrm{CTI}_P$ was found by minimizing the $\chi^2$ (Equation \ref{eq:chi2}) over flat field signal levels below 5000 electrons, given data $D(x, S_F)$ with error $\sigma(x, S_F)$.  To explore potential temperature dependence of the emission time constant, the minimization procedure was repeated for a second set of flat field images taken at \ang{-90} C.

\begin{equation}\label{eq:chi2}
\sum_{S_F} \chi^2_{S_F} = \sum_{S_F} \sum_x \frac{D(x, S_F) - S_O(x, S_F)}{\sigma(x, S_F)^2}.
\end{equation}

\begin{figure}[htb]
\begin{center}
\begin{tabular}{c}
\includegraphics[height=8cm]{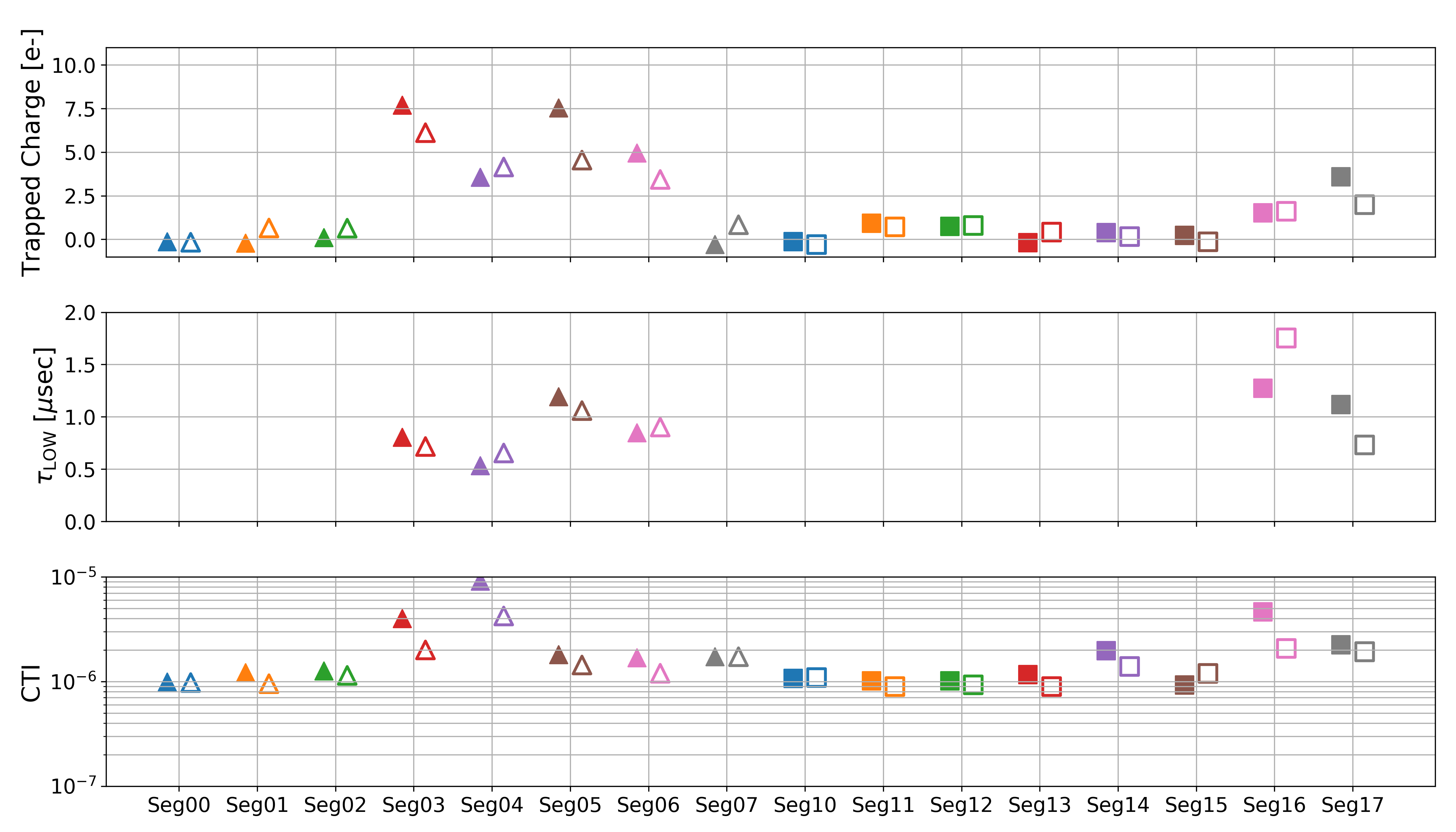}
\end{tabular}
\end{center}
\caption 
{ \label{fig:lowflux}
Exponential tail fitting results, per segment, jointly fit to EPER measurements for signals below 5,000 electrons for $\ang{-100}$ C (Solid Markers) and $\ang{-90}$ C (Outline Only Markers) sensor operating temperatures.   Parameter estimate errors of $\pm$ 5 electrons, $\pm$ 0.5 $\mu$sec and $\pm$ 10\% relative error  respectively are not shown for visual clarity.} 
\end{figure}

The results of the exponential tail fitting are shown in Figure \ref{fig:lowflux}.  The low signal inefficient segments are identifiable by the trapping of greater than 1 electron (calculated by summation of the fitted exponential contribution for overscan pixels $x \geq 1$) and exhibit emission decay time constants between 0.5 and 1.5 $\mu$sec, though no substantial temperature dependence is observed.  It is known that fitting a curve with multiple exponentials can give an unstable fit result, and does not take into account charge recapture in the overscan pixel tail.\cite{Skottfelt2018}  For these reasons, the results of this minimization are not considered precision measurements, but an estimation of the emission time constant can be illuminating for identifying possible trapping species.  A more careful treatment of the exponential tail fitting to determine emission time constants is a subject of future work.

The calculated emission time constants of between 1 $\mu$sec and 0.5 $\mu$sec do not correspond strongly with known defects for n-channel CCDs at temperatures of \ang{-100} C, though emission time constants are believed to occupy a broad distribution of values.\cite{Skottfelt2018}  The three most likely candidates for a known trap species are Silicon-Arsenic traps, divacancy traps, and Carbon-interstitial-Phosphorus-substitution traps, which have emission time constants of O($10^{-6}$) seconds, O($10^{-5}$) seconds and O($10^{-5}$) seconds respectively at $\ang{-100}$ C.

Another possibility for charge trapping beyond radiation or bulk traps would be a process or design trap in the serial register.  Due to the segment geometry (Figure \ref{fig:itl_layout}), all serial registers have an additional 3 serial prescan pixels that angle away from the pixel region, in order to allow sufficient space for the output and reset amplifiers.  If a process trap were to occur in the serial prescan pixels, localized trapping could occur during transfer through these pixels.  For this scenario, the emission time constant may not have a strong dependence on sensor temperature.

\subsection{\texorpdfstring{\textsuperscript{55}Fe}{Fe55} Profile measurements}
\label{sect:fe55}

Another common method for determining the CTI is to use \textsuperscript{55}Fe soft x-rays that have a known charge deposition in silicon CCDs.  CTI is calculated by analyzing how the total x-ray hit flux decreases as a function of the number of transfers \cite{Yates2017}.  For the case of deep-depletion CCDs such as the ITL sensors studied, charge diffusion of greater than 3.6 $\mu$m in the 100 $\mu$m thick silicon causes the deposited electrons to spread into adjacent pixels, complicating the precision measurement of CTI \cite{Yates2017}.  Results of measuring CTI using \textsuperscript{55}Fe central pixel flux and integrated flux (using the methods outlined by Yates, Kotov and Nomerotski)\cite{Yates2017} were largely inconclusive; an additional complication to these measurements is the smaller number of serial transfers (512 transfers) compared to parallel transfers (2000 transfers) from which to measure a slope (electrons/transfer) in the \textsuperscript{55}Fe x-ray fluxes.

\begin{figure}[htb]
\begin{center}
\begin{tabular}{c}
\includegraphics[height=10cm]{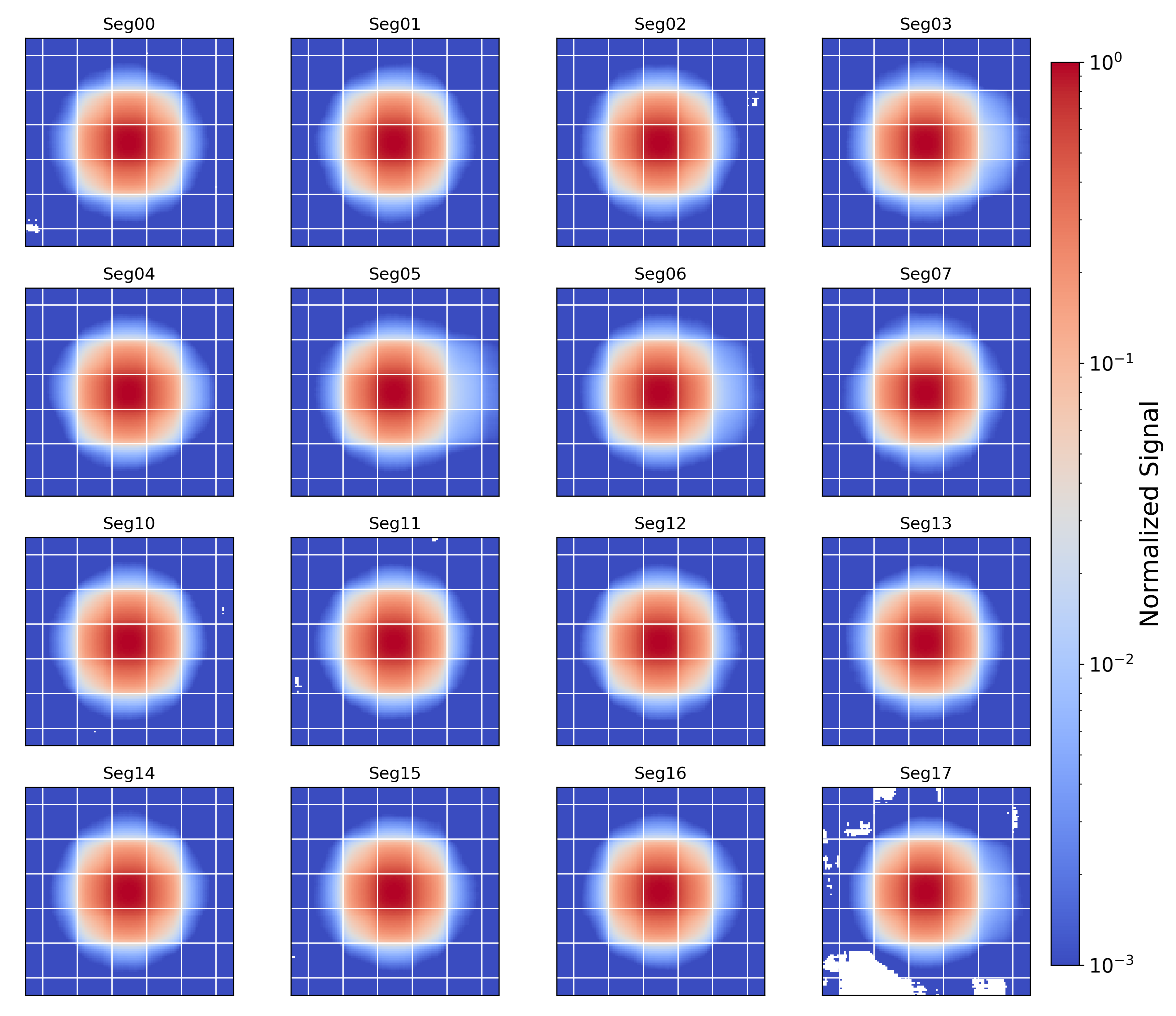}
\end{tabular}
\end{center}
\caption 
{ \label{fig:fe55}
Normalized aggregated \textsuperscript{55}Fe profile for each CCD segment (grid lines indicate pixel boundaries).  High CTI is identifiable by a left-right asymmetry in pixel signal at the 1\% level.} 
\end{figure}

The large number of \textsuperscript{55}Fe x-ray acquisitions taken during sensor gain stability testing can be used to demonstrate the deferred charge effect on low signal sources in the low signal inefficient segments.  An aggregated \textsuperscript{55}Fe x-ray footprint for each amplifier of the CCD was created by re-sizing each individual footprint (totalling over 10000 per segment) via oversampling and stacking the footprints by their sub-pixel centroid, calculated using the Sloan Digital Sky Survey adaptive moments shape algorithm.\cite{Bernstein2002}  The effect of the large deferred charge at low signal is identifiable in specific segments as a 1\% left-right asymmetry in the normalized \textsuperscript{55}Fe profile (Figure \ref{fig:fe55})

\section{High Signal Deferred Charge}
\label{sect:highsignal}

\begin{figure}[htb]
\begin{center}
\begin{tabular}{c}
\includegraphics[height=10cm]{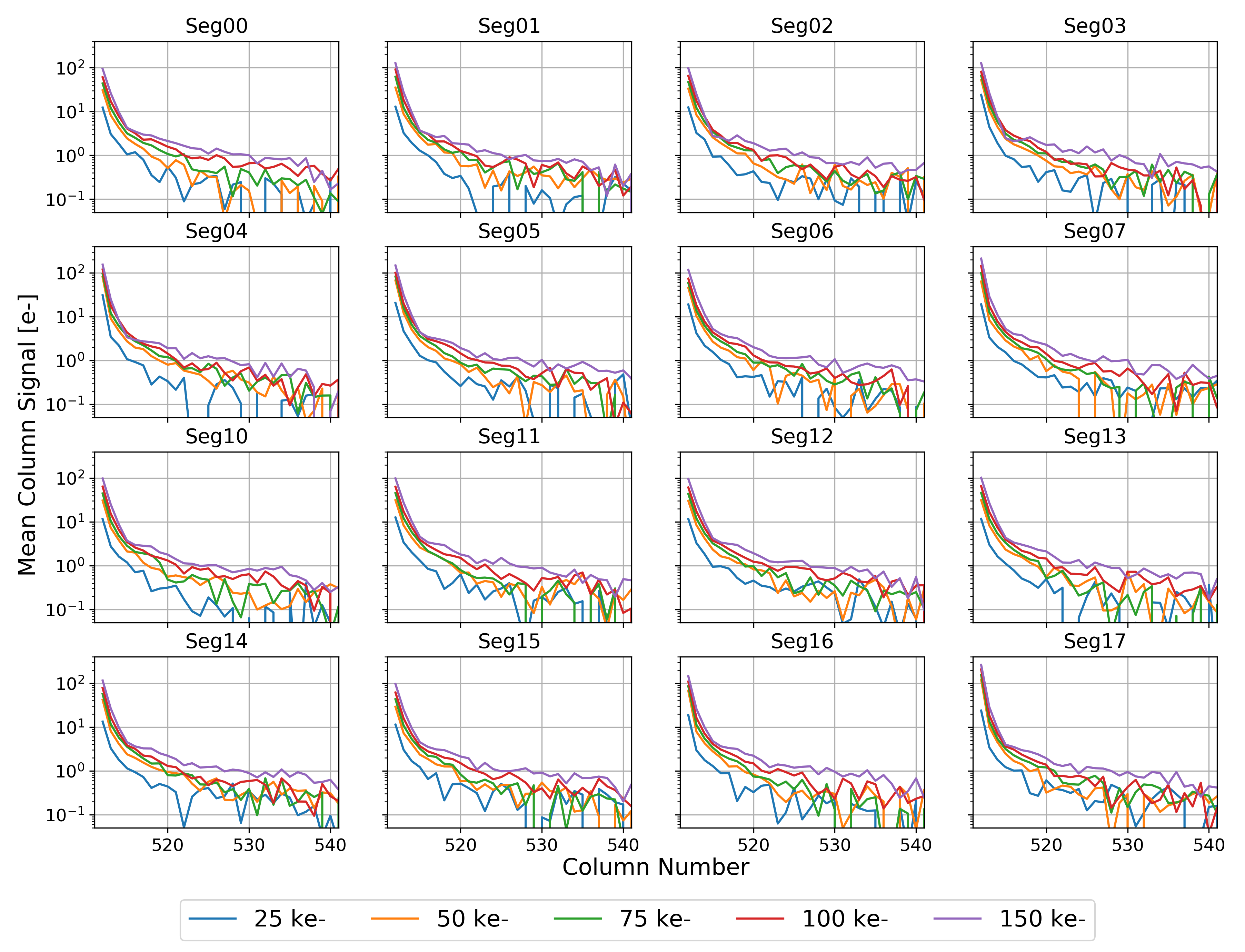}
\end{tabular}
\end{center}
\caption 
{ \label{fig:eper_high}
Mean signal in overscan columns after a flat field image (plotted on a log scale), for five different high signal levels, for each CCD segment.  All segments show deferred charge extending many pixels into the overscan pixel region. Error on the mean column signal is 0.14 electrons.} 
\end{figure}

The mean signal in overscan columns as a function of column number for flat field signals greater than 25,000 electrons is shown in Figure \ref{fig:eper_high}.  All segments show a deferred charge exponential tail extending many pixels into the serial overscan region that increases with increasing flat field signal.  These exponential tails do not have a strong effect on the measured CTI from EPER (Figure {\ref{fig:cti}}) due to the fact that only first overscan two pixels were used during those calculations.

\begin{figure}[htb]
\begin{center}
\begin{tabular}{c}
\includegraphics[height=8cm]{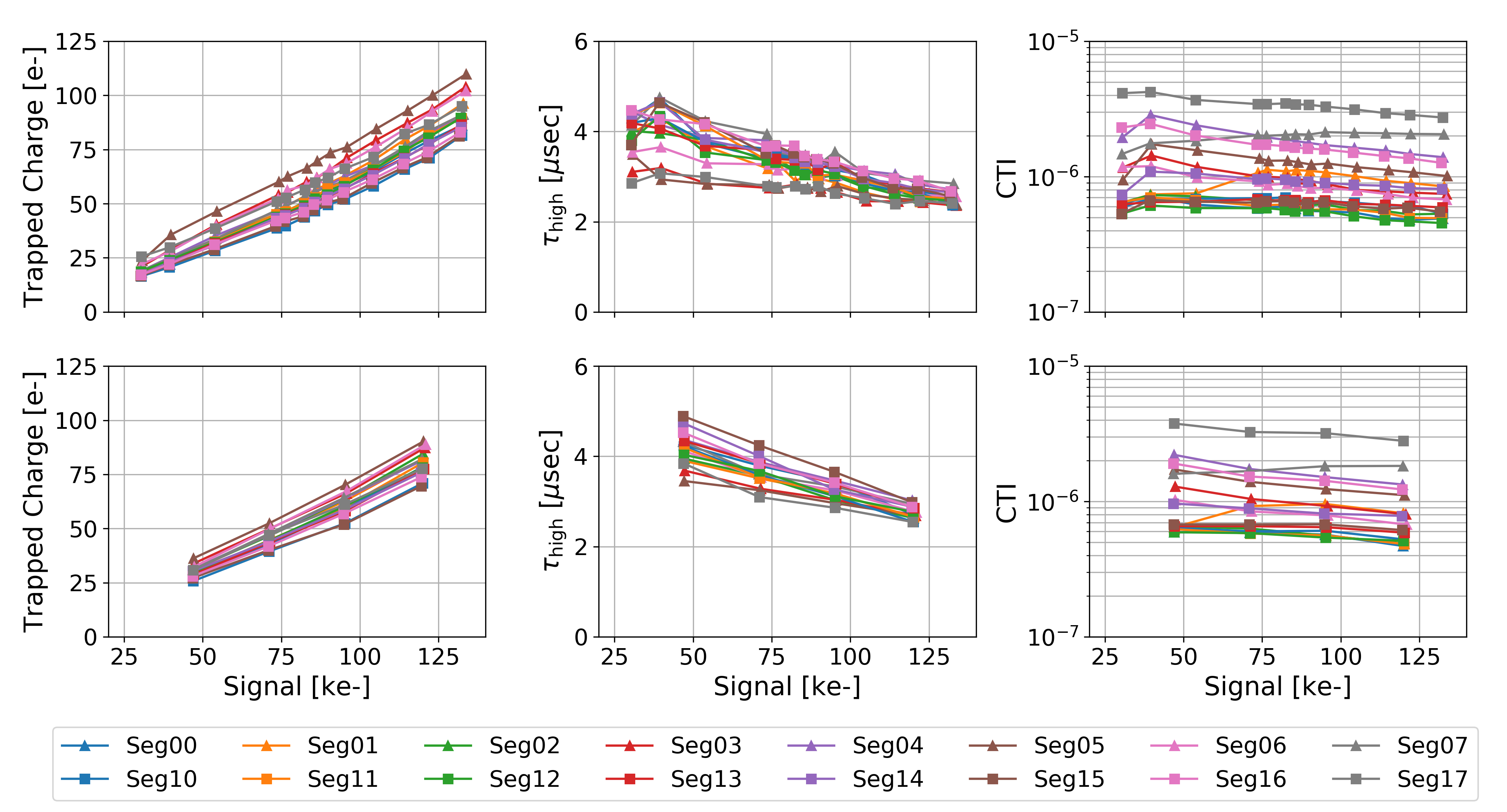}
\end{tabular}
\end{center}
\caption 
{ \label{fig:highflux}
Exponential tail fitting results, per segment, for high signal EPER measurements (greater than 25,000 electrons), as a function of flat field signal for $\ang{-100}$ C (Top Row) and $\ang{-90}$ C (Bottom Row) sensor operating temperatures.  Parameter estimate errors of $\pm$ 5 electrons, $\pm$ 0.5 $\mu$sec and $\pm$ 10\% relative error  respectively are not shown for visual clarity.} 
\end{figure}

The exponential tail fitting procedure described in Section \ref{sect:lowsignal} was repeated for the mean overscan column signals using flat field images with signals greater than 25,000 electrons.  However, as the total deferred charge appears to increase with increasing signal, the $\chi^2$ minimization was performed separately for each flat field signal level.  The results for the flat field data set taken at \ang{-100} C and \ang{-90} C are shown in Figure \ref{fig:highflux}.  All segments show the same slowly varying trends; the amount of trapped charge increases roughly linearly with increasing signal while emission time constant and proportional CTI loss per transfer decrease  with increasing signal.  As was the case for low signal levels, more sophisticated exponential tail fitting is a subject for future work.

The minimization results indicate trapping of up to 100 electrons or more, in all segments, which would indicate a large number of traps in the serial register.  All segments pass the more stringent parallel CTI specification of less than $3 \times 10^{-6}$ and show no evidence for a large number of traps throughout the bulk silicon; routine measurements of CTI at the $10^{-7}$ level are common for parallel CTI at both high and low signal levels.  A more likely explanation for the measured deferred charge is a hysteresis (or drift) of the electronic offset level, whereby the electronic offset has increased during the readout of the preceding high signal pixels and decays back to its nominal value over a timescale larger than the single pixel read time, which has not been completely removed by the correlated double sampling of the segment video signals.

\section{Conclusions}
\label{sec:conclusions}

The measurements of deferred charge described in Section \ref{sect:lowsignal} and Section \ref{sect:highsignal} indicate that in addition to proportional CTI loss that occurs at each pixel transfer, ITL CCDs have additional contributions to the total deferred charge measured in serial overscan pixels.  A subset of segments have large CTI (measured from EPER) likely due to trapping in the serial register of O(10) electrons, that saturates at signal levels of under 100 electrons (though some segments may not saturate till signal levels of several 100 electrons) resulting in fixed loss during serial readout.  All segments exhibit increased proportional CTI loss of up to 100 electrons at high signal levels, resulting in large deferred charge exponential tails in the serial overscans.  Due to the large amount of deferred charge, it is more likely that this is due to an electronic offset drift effect rather than a large additional population of traps in the serial register, separate from the low signal effect.

The results presented in this paper are only for a single sensor, however all the described measurements and analyses have been performed for a large number of LSST ITL CCDs, with similar results.  All segments on all the ITL CCDs tested exhibit the high signal deferred charge effect.  The large majority of ITL CCDs tested have a subset of segments that exhibit the low signal deferred charge effect, and there is no correlation between segment location on the CCD and the propensity for the segment to be low signal inefficient.

The impact of the deferred charge effects outlined in this paper on astrophysical images is largely unknown.  The deferred charge effect in low signal inefficient segments can be measured using \textsuperscript{55}Fe x-ray hits, as a less than 1\% asymmetry in the hit profiles.  The deferred charge effect at high signal levels on bright sources has not been measured, but will be the subject of study during full camera focal plan integration and testing. Of particular interest is the potential for deferred charge to affect the focal plane point spread function characterization, source moment measurements, and astrometry calculations.  A more careful treatment of the exponential tail fitting procedure at all signal levels up to full-well can be used to develop a model of the deferred effects to estimate the impact of the serial deferred charge on these measurements; this is an ongoing research focus.

\acknowledgments 

This material is based upon work supported in part by the National Science Foundation through Cooperative Agreement 1258333 managed by the Association of Universities for Research in Astronomy (AURA), and the Department of Energy under Contract No. DE-AC02-76SF00515 with the SLAC National Accelerator Laboratory. Additional LSST funding comes from private donations, grants to universities, and in-kind support from LSSTC Institutional Members.

We thank Stuart Marshall and Yousuke Utsumi for their assistance in the acquisition of LSST sensor data.  Thanks is also given to Pierre Antilogus and Claire Juramy for their ongoing work in LSST CCD optimization and related studies of  deferred charge.


\bibliography{ispa}   
\bibliographystyle{spiejour}   


\vspace{2ex}\noindent\textbf{Adam Snyder} is an graduate student at Stanford University. He received his BS in physics from the University of Illinois at Urbana-Champaign in 2014.  His current research interests are characterization of imaging sensors for precision optical and near-infrared astronomy and cosmological measurements of dark energy and dark matter. He is a member of SPIE.

\vspace{1ex}
\noindent Biographies and photographs of the other authors are not available.

\listoffigures

\end{spacing}
\end{document}